\documentclass[
aps,
reprint,
groupedaddress,
amsmath,
amssymb,
nofootinbib]{revtex4-2}

\usepackage{graphicx}
\usepackage{dcolumn}
\usepackage{bm}
\usepackage{hyperref}
\usepackage{multirow}
\usepackage{tabularx} 
\usepackage{enumitem} 
\usepackage{lineno}

\begin{document}

\preprint{arXiv:2001.07277}

\title{Positive Impacts on Student Self-Efficacy from an Introductory Physics for Life Science Course Using the Team-Based Learning Pedagogy}

\author{Brokk Toggerson}
 \email[Correspondence can be directed to ]{toggerson@physics.umass.edu}
\author{Samyukta Krishnamurthy}
\author{Emily E. Hansen}
\affiliation{Physics Department, University of Massachusetts Amherst, Amherst, Massachusetts 01002, USA}

\author{Chasya Church}
\affiliation{Nassau Re, Hartford, Connecticut 06103, USA}

\date{\today}

\begin{abstract}
We present the impact on student self-efficacy of an introductory physics for life-science students course taught using a Team-Based Learning pedagogy. We measured self-efficacy using the validated quantitative Sources of Self-Efficacy in Science Courses – Physics (SOSESC-P) survey developed by Fencl and Scheel. Data were collected both at the beginning and end of the semester to evaluate the impact of shifts in individual self-efficacy. After describing the key features of the pedagogy, we find that the Team-Based Learning system at University of Massachusetts Amherst, results in significant improvements for she-identifying individuals from three of the four sources of self-efficacy identified by Bandura as well as in three of four investigated attributes of the course. We also investigated the predictive power of self-efficacy on individual student performance using logarithmic regression. For our course, the shift in self-efficacy between the beginning and end of the semester is more important than a student's pronouns in predicting attaining at least a B on individual assignments.
\end{abstract}

\maketitle


\section{\label{sec:Introduction} Introduction}

Over the past several years, there has been a growing recognition of the important role of student-self efficacy in the introductory physics classroom. Self-efficacy is a concept originally defined by Bandura as, ``beliefs in one's capabilities to organize and execute the courses of action required to produce given attainments'' ~\cite{Bandura}. Research suggests that students with high self-efficacy will be more likely to persist in the face of struggle and therefore will be more likely to remain and succeed in particular fields ~\cite{House1}~\cite{House2}~\cite{Lent}. Therefore, measuring the self-efficacy impact of various pedagogies, such as Modeling Instruction~\cite{Dou} and Peer Instruction~\cite{Miller}, as well as identifying experiences which improve self-efficacy~\cite{Identifying-SE-MI} have been areas of particular interest. 

During this same period, and in parallel, there has been an increased interest in the teaching of introductory physics for life science (IPLS) students: a group of students with traditionally low interest in physics~\cite{Hall}. Traditionally, courses for life-science students have covered similar content to those for engineering and physics majors. However, motivated by reports such as the \textit{Scientific Foundations for Future Physicians}~\cite{Future-Physicians}, which identify the growing importance of physical understanding to the life-sciences, groups such as the NEXUS project have begun to critically reexamine the content of these courses~\cite{NEXUS}.   

In the present study, we are interested in the self-efficacy impacts of a large-enrollment IPLS course taught at University of Massachusetts Amherst using a Team-Based Learning (TBL) pedagogy based upon the work of Michaelsen \textit{et al.}~\cite{Michaelsen}. The impacts of this particular teaching strategy on self-efficacy in large-enrollment physics courses are not well documented in the literature. After a brief discussion of our particular course, we explore shifts in student self-efficacy as measured by the validated Sources of Self-Efficacy in Science Courses Survey for Physics (SOSESC-P) developed by Fencl and Scheel~\cite{Fencl} with a particular focus on the impacts dis-aggregated by student pronoun identification. For while the correlations between gender and performance in introductory physics are well known~\cite{Kost}, and the work of Eddy \textit{et al.} demonstrates that inequities persist in she-identifying dominated introductory biology courses, the relationships between pronoun identification and performance in an IPLS course are less well explored. Finally, we then check self-efficacy's power as a predictor for student success in the course using a logistic regression, concluding with some thoughts on impacts for future instruction.

\section{First-Semester IPLS at University of Massachusetts Amherst} \label{sec:UMassIplsTbl}

The first of the two-semester IPLS sequence at University of Massachusetts Amherst has six sections of 100 students each and is taught in a studio-style room with eleven round tables that can each seat up to ten students. Each table in the room has electrical outlets and laptops for laboratory activities as well as dedicated whiteboard space on a nearby wall. There are a total of six sections with 100 students each. Similar to other studio-based active-learning collaborative systems such as SCALE-UP~\cite{SCALEUP} and Collaborative Problem Solving~\cite{CPS}, students in the TBL pedagogy spend the majority of class time working in teams to solve problems. The curriculum is inspired by other IPLS courses such as the NEXUS project~\cite{NEXUS}, while being tailored to our student population which has a significant proportion of kinesiology students. The result is a five unit course outlined in Table~\ref{tbl:units}. To our knowledge, University of Massachusetts Amherst is the only institution to be implementing an IPLS curriculum, using the TBL pedagogy, at this scale.

\begin{table}
\nolinenumbers
\caption{\label{tbl:units} The five units of the first semester IPLS course at University of Massachusetts Amherst. The timing of each exam is also indicated.} 
\begin{tabularx}{\columnwidth}{l r@{\hskip 1em}X}
\hline \hline
  & Days & Topics \\
  \hline
Unit 1   & 7 & Kinematic concepts and introduction to Newton's Laws in one-dimension. \\
Unit 2  & 9 & Newton's Laws with multiple forces in multiple dimensions. \\
\multicolumn{3}{c}{Exam I} \\
Unit 3 & 4 & Static torque with an emphasis on bio-mechanics. \\
Unit 4 & 8 & Conservation of energy with an emphasis on developing a trans-disciplinary picture across distance scales. \\
\multicolumn{3}{c}{Exam II} \\
Unit 5  & 6 & Statistical interpretation of entropy. \\
\multicolumn{3}{c}{Exam III (During finals period)} \\
\hline \hline
\end{tabularx}
\end{table}

\begin{figure}[tbh!]
    \includegraphics[width=\columnwidth]{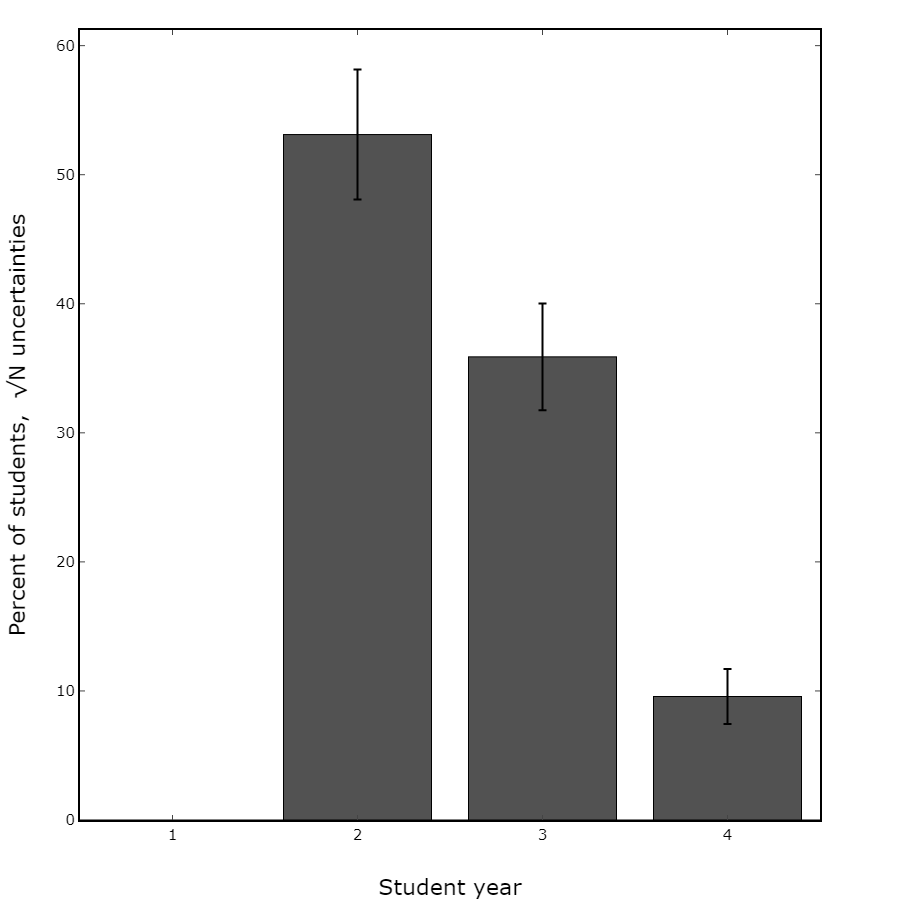}
    \caption{\label{fig:year} The year distribution for the student population for three sections of first-semester TBL IPLS at University of Massachusetts Amherst ($N = 206$).}
\end{figure}

\begin{figure}[tbh!]
    \includegraphics[width=\columnwidth]{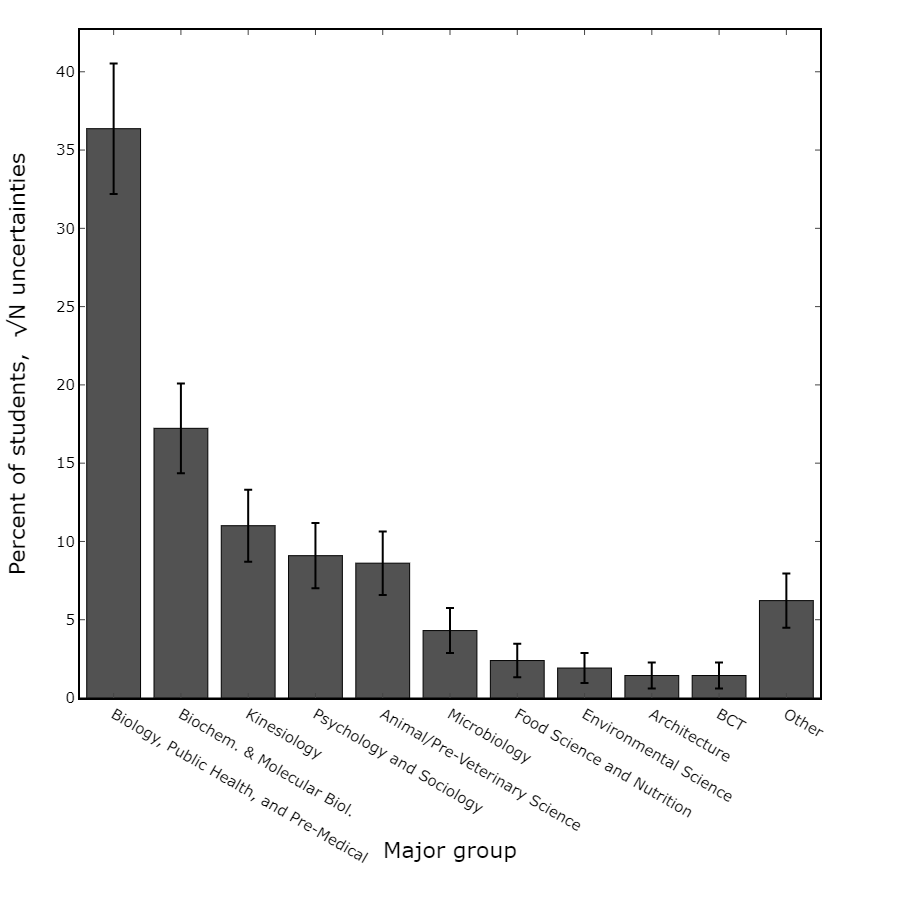}
    \caption{\label{fig:major} The distribution of student majors for three sections of first-semester TBL IPLS at University of Massachusetts Amherst ($N = 206$).}
\end{figure}

A total of 66.5\% of our student population identify with ``She/Her'' pronouns, and 32.5\% identify with ``He/Him.” The remaining 1.5\% identify some other way. This She/He split is consistent with national trends in the life sciences~\cite{Eddy}. Similar to many other IPLS courses, the population is also predominately second- and third-year students as seen in Figure~\ref{fig:year}. The year demographic, however, is strongly correlated with student major, the distribution of which can be seen in Figure~\ref{fig:major}. For example, Biology students tend to take physics in their second year while Microbiology students tend to take it in their third. The lack of first-year students is due to the fact that this population spends their first year typically taking calculus, biology, and chemistry. While the course population is clearly dominated by life-science students, there are other majors in the course, notably Architecture and Building and Construction Technology (BCT). In addition, a few other students who choose to take this course to fill a physical-science general-education requirement are present.

\subsection{\label{sec:TBL} Overview of the Team-Based Learning Pedagogy}

The team-based learning pedagogy used in the first semester IPLS course at University of Massachusetts Amherst is a flipped model with significant individual accountability. Before the first day of each unit outlined in Table~\ref{tbl:units}, students are responsible for completing readings and preparatory homework assignments. In our course, the average unit’s preparatory assignment comprises sixty pages of reading and about twenty problems. These problems would be typically classified as \textit{Remember} or \textit{Understand} under the revised Bloom's Taxonomy of Krathwohl~\cite{RevisedBlooms}. Some example problems can be seen in Appendix~\ref{app:problems}.

After completing the homework, students then on the first day of the unit undergo what Michaelsen \textit{et al.} call the \textit{Readiness Assessment Process}: taking a ten-question multiple choice quiz based upon their homework. They take this quiz first individually and then with their teams~\cite{iRAT}. The only exception to this pattern is for the first unit. For the first unit, based upon the work of Miller \textit{et al.} which suggests that summative assessments too early in the course can be detrimental to self-efficacy, there is no quiz. Instead, students are provided an ungraded sample quiz. The remaining days of each unit are then spent on in-class problem solving at the whiteboards, conceptual multiple choice questions similar to the \textit{ConcepTests} of Peer Instruction~\cite{PI}, and laboratory activities. 

After Units 2, 4, and 5, students take exams. All exams are cumulative. Exams I and II are during the semester while Exam III is during the finals period. The custom at University of Massachusetts Amherst is to have exams outside of class time, typically a Tuesday or Thursday night, with the next unit's homework due the following Monday. Following the standard TBL pedagogy, students first take the Exam Individually and then as a team. The work of Heller with the Collaborative Problem Solving pedagogy indicates that students need approximately 20 minutes to solve a context rich problem on an exam~\cite{CPS}. In order to ensure that the exams are doable in one-hour, they comprise 10 multiple choice questions and two long-answer. In the interests of equity and to minimize the need for accommodations for those students with disabilities, everyone is then given two hours to complete the exam. 

\subsection{\label{sec:tbl-teams} Teams in TBL}

Team size and composition represent two of the most significant differences between TBL and other, similar, active-learning collaborative pedagogies. Moreover, we suspect these differences may have an important impact on self-efficacy. Many other systems, such as Collaborative Problem Solving~\cite{CPS} and SCALE-UP~\cite{SCALEUP}, typically use groups of three built by the instructor with specific group roles for each student. These groups are then typically shuffled a few times a semester. Others, such as Peer Instruction~\cite{PI}, use ad-hoc groups of two-to-three students that are self-organized based upon proximity in the learning environment. For the TBL pedagogy, in contrast, students are organized into larger groups of five, remain in their teams for the entire semester, and self-organize their roles. 

The work of Kowitz and Knutson suggest that, for sufficiently difficult tasks, groups of five to seven are optimal~\cite{Kowitz-Knutson}. These larger, five-person teams have the benefit of a wider, more diverse pool of knowledge and skills available to the team~\cite{Michaelsen}. As has been documented in the literature, more diverse teams are generally more successful at tackling challenging problems~\cite{Rock}~\cite{Horwitz}.  In addition to larger groups having, almost by definition, a larger variety of perspectives, the teams are constructed to be diverse using the CATME system~\cite{CATME-teamformation}. This system ensures that the groups are heterogeneous across a variety of dimensions including for example major, GPA, year, pronoun identification, and preferred leadership style. Simultaneously, team construction strives to minimize the potentially detrimental effects of soloing~\cite{soloing} by ensuring that those students who are typically underrepresented in physics are not in teams by themselves. 

In addition to the larger team size, the teams in TBL exist for longer than in many other active learning pedagogies, where teams are shuffled a couple of times a semester. This longer duration is important to provide teams the 20-25 hours, about five-to-six weeks of class time, to come together and learn each others' strengths and weaknesses~\cite{Watson}. Along the way, students figure out their own group dynamics, eliminating the need for prescribed ``roles,'' which serve the useful function of expediting group cohesion in impermanent groups~\cite{Scheidel-Crowell}. With this amount of time, as Michaelsen says, ``groups develop into effective self-managed learning teams.''~\cite{Michaelsen}

Another important characteristic of TBL is the emphasis placed on evaluating the work of teams as measuring the performance of the team results in more cooperation and better team performance~\cite{Hackman}. At University of Massachusetts Amherst, 35\% of the total grade is team-based, which is in line with the recommendations of Michealsen \textit{et al}~\cite{Michaelsen}. As shown in Table~\ref{tbl:grade-breakdown}, this team-component includes both the collaborative portions of quizzes and exams as well as laboratory exercises. In comparison, other pedagogies with organized teams, such as Collaborative Problem Solving and SCALE-UP, connect a small fraction of the grade to group work: 10\% in the case of Collaborative Problem Solving~\cite{CPS}. Meanwhile, strategies which use informal groups, such as Peer Instruction, typically do not grade on team performance~\cite{PI}. 

\begin{table}[tbh!]
\caption{\label{tbl:grade-breakdown} The grade breakdown for the first-semester IPLS course at University of Massachusetts Amherst.}
\begin{tabular}{l r| l r}
\hline \hline
\multicolumn{2}{l|}{Individual Components} & \multicolumn{2}{l}{\hspace{0.5em} Team Components} \\
\hline
Individual Exam I & 15\% \hspace{1em} & \hspace{0.5em} Team Exam I & 5\% \\
Individual Exam II & 15\% \hspace{1em} & \hspace{0.5em} Team Exam II & 5\% \\
Individual Exam III & 15\% \hspace{1em} & \hspace{0.5em} Team Exam III & 5\% \\
\hline
Inidividual Quizzes & 10\% \hspace{1em} & \hspace{0.5em} Team Quizzes & 10\% \\
\hline
Preparatory Homework & 10\% \hspace{1em} & \hspace{0.5em} Laboratory & 10\% \\
\hline
Individual Total & 65\% \hspace{1em} & \hspace{0.5em} Team Total & 35\% \\
\hline \hline
\end{tabular}  
\end{table}

\begin{figure}[tbh!]
	\includegraphics[width=\columnwidth]{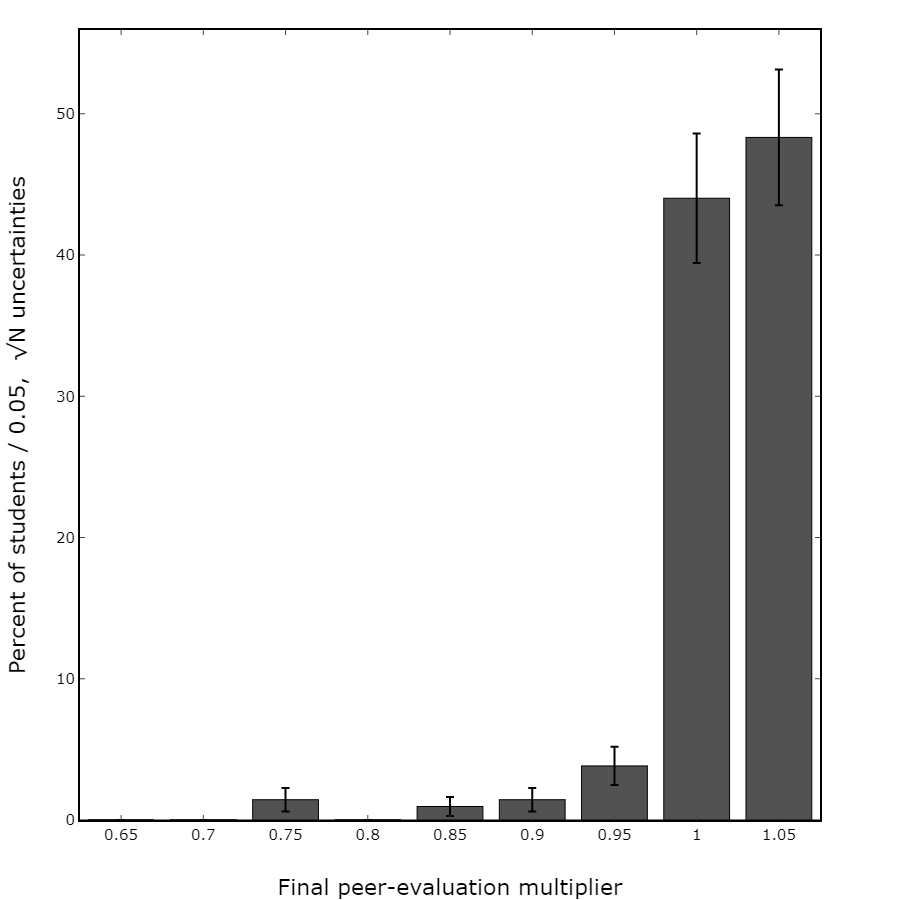}
	\caption{\label{fig:peerMultiplier} The distribution of final peer evaluation scores for three sections of a first-semester IPLS TBL course at UMass-Amherst $N = 206$. This score is multiplied to all team grade components. Clearly, most students earn a 1.0 or higher.} 
\end{figure}

To allay student fears about ``slackers'' on their teams and to reward those students who go above-and-beyond in helping their team-mates learn, there must be a method of evaluating individual contributions to the team~\cite{Schnake-1991}. At University of Massachusetts Amherst, we use a multiplier-based peer-evaluation method following Fink's in Appendix B of \cite{Michaelsen}. Twice a semester, students complete an evaluation of their teammates. The first evaluation is purely formative and occurs mid-semester. The second evaluation, conducted during the finals period, results in each student receiving a multiplier which is then multiplied to all team grades. Scores up to 1.05 are possible, giving a boost to those students who, in the eyes of their peers, were instrumental to their team's success. As visible in Figure~\ref{fig:peerMultiplier}, most students earn a multiplier of 1.0 or above, with few earning less than a 0.95. The instructor, of course, reviews these scores to check for biases or to take into account specific mitigating circumstances.

\section{\label{sec:method} Methods}

Three different instructors are involved in teaching the six sections of first-semester IPLS at University of Massachusetts Amherst. All use variations of the curriculum described in Section~\ref{sec:UMassIplsTbl}\footnote{This research methodology was approved by the University of Massachusetts Amherst IRB. Protocol number 2018-4975.}. Therefore, no control lecture section was available. In order to eliminate variables arising from different instructors and slight variations in the curricula, the potential participant pool for this research is from three sections taught by the author, Toggerson, during the Fall 2018 semester. These three sections total 260 students. The three main sources of data are: grade information; the Sources of Self-Efficacy in Science Courses Survey - Physics (SOSESC-P) developed by Fencl and Scheel~\cite{Fencl}; and the CATME peer evaluation software~\cite{Loughry}~\cite{Ohland}~\cite{Loignon}. Our research followed a blind-analysis paradigm: the entire analysis plan and chain was developed on a small set of students from a previous semester.

\subsection{\label{sec:SOSESC-P} The Sources of Self-Efficacy in Science Courses Survey - Physics (SOSESC-P)}

The SOSESC-P is a validated 33-question survey that looks at each of the four sources of self-efficacy described by Bandura: mastery experiences, vicarious learning, verbal persuasion, and physiological state. Mastery experiences are indicated by Bandura as being ``the most influential source of efficacy information because they provide the most authentic evidence of whether one can muster whatever it takes to succeed''~\cite{Bandura}. Meanwhile, vicarious learning plays a related role wherein students define their success not through some personal or absolute standard, but by comparison to their peers. Such comparisons will, by default, play an important role in a physics course during the period prior to summative assessments. In contrast, verbal persuasion ``may be limited in its power to create enduring increases in efficacy, but it can bolster self-change if the positive appraisal is within realistic bounds''~\cite{Bandura}. Finally, a person's physiological or emotional state can impact self-efficacy; the fear of being incapable can be sufficiently distracting as to produce the very inadequacy which was feared.

In addition, the survey also looks at these four sources in different attributes of a physics course: attainment (getting good grades), understanding, attentiveness, test taking, and recall and recognition. The SOSESC-P was administered at the same time as the Colorado Learning Attitudes towards Science Survey (CLASS)~\cite{CLASS}. In order to keep the present study manageable and focused, the data from the CLASS will be presented in future work. The combined survey was administered twice during the semester. Our Institute for Teaching Excellence and Faculty Development administered the surveys to remove the possibility of conflicts-of-interest arising from instructors administering research surveys in their own courses. The first administration was during the first two weeks of the course to gather data about students' incoming self-efficacy beliefs. The second administration began during the final week of class and continued through the finals period. Following IRB 1055924 at University of Central Florida, students were given extra credit for completing both surveys, but were not required to consent to their data being used for research purposes to earn that credit. 


For a student's results to be considered valid, their responses had to pass a series of quality checks. Students were required to spend over three minutes on the SOSESC-P portion of the survey and skip no more than three questions. In addition, students were required to correctly answer question 31 on the CLASS, ``We use this statement to discard the survey of people who are not reading the questions. Please select agree-option 4 (not strongly agree) for this question to preserve your answers.'' Those surveys which did not pass these checks were discarded, amounting to 36.4\% of surveys. Even if a student's survey data were discarded, their grade data, however, were still used to gain a more holistic picture of the culture of the course.

\subsection{\label{sec:CATME} The CATME Peer Evaluation}

In addition to the SOSESC-P survey, significant insights came from consenting students' responses to the required end-of-semester peer evaluation which opened during the last day of class and was available for the week of the finals period. As described in Section~\ref{sec:tbl-teams}, students in the TBL pedagogy evaluate their peers, resulting in a multiplier applied to the 35\% of their grade determined by team assignments. At University of Massachusetts Amherst, this peer evaluation is accomplished using the CATME peer evaluation system. In this research-based system, students rate their peers, and themselves, on a number of dimensions including ``Contributing to the team's work,'' ``Interacting with teammates,'' ``Keeping the team on track,'' and ``Expecting quality''~\cite{Ohland}. The results of this peer-evaluation were used to gain insight into the effectiveness and cohesion of teams in the course.   

\section{\label{sec:results} Results}

Of the 260 students in the course, 206 consented for their data to be used in the present study (79.2\%). As previously mentioned, 36.4\% did not pass the quality checks leaving the survey results from 131 students. In this sub-population, 91 individuals identify as `She,' 37 as `He,' and 3 use some other pronoun identification; a gender ratio consistent with the larger 206-student group who consented to participate. 

A non-parametric Wilks’ lambda test helped ensure that the surveys which passed all quality checks were not from a demographically distinct subset in terms of pronouns, majors, and incoming GPAs relative to the broader sample of consenting students. The multivariate Wilks’ lambda test, detailed in~\cite{wilks}, results in a test statistic $F_{LBH}$ which characterizes the variation within versus between orthogonal groups. The test is specifically designed to avoid continuity and normality assumptions, allowing it to be used for ordinal and categorical demographic data. In this case, the two groups are the 131 students whose surveys passed all checks and the 75 who did not. After calculating $F_{LBH}$ for these two groups, the test statistic was calculated for 1000 random pairs of groups of 131 and 75 students each. The results of this Monte Carlo simulation indicate that if the 206 students were divided randomly into a group of 131 and 75, those groups would be more demographically distinct than groups of students whose surveys passed quality checks versus those whose did not 41\% of the time. 

The pre- and post-scores for overall self-efficacy, as measured by the SOSESC-P, are broken down by pronoun preference in table~\ref{tbl:SE-results}. The Cohen's $d$ of approximately 0.2 indicates a small positive shift for both the class as a whole and for `she`-identifying individuals. For the people who identify as `he,' the fact that the 95\% CL for Cohen's $d$ crosses zero is indicative of limited statistics. To place these results into context, a study done by Sawtelle \textit{et al.}~\cite{FIU-2010CP} showed that students in a lecture-based course at a large public university showed a statistically significant decrease in self-efficacy for all students. Similarly, Fencl and Scheel~\cite{Fencl} also showed that more active learning environments were correlated with increases in self-efficacy.

\begin{table*}[tbh!]
\caption{\label{tbl:SE-results} A summary of the self-efficacy as measured by the SOSESC-P in our course, disaggregated by gender. Note, the `All' does not add to `She' plus `He' due to three students who identify with some other pronouns.}
\begin{tabular}{l r r r}
\hline \hline
                      &  All                           & She                            &  He \\
\hline
$N$                & $131$                        & $91$                          & $37$ \\
 pre-               & $3.419$                     & $3.375$                      & $3.542$ \\
 post-              & $3.566$                     & $3.528$                     & $3.662$ \\
 $t$                 & $3.862$                     & $3.355$                      & $1.716$ \\
 $p$                 & \hspace{3em} $1.8 \times 10^{-4}$ & \hspace{3em}$1.2 \times 10^{-3}$ & \hspace{3em} $9.5 \times 10^{-2}$ \\
 Cohen's $d$     & $0.286$                    & $0.322$                    & $0.203$ \\
 95\% CI Cohen's $d$ (UL, LL) & $(0.517, 0.162)$ & $(0.563, 0.139)$  & $(0.609, -0.049)$ \\
 \hline \hline
\end{tabular}
\end{table*}

\subsection{\label{sec:results-se-sources} Sources of Self-Efficacy}

\begin{figure}[tbh!]%
	\includegraphics[width=\columnwidth]{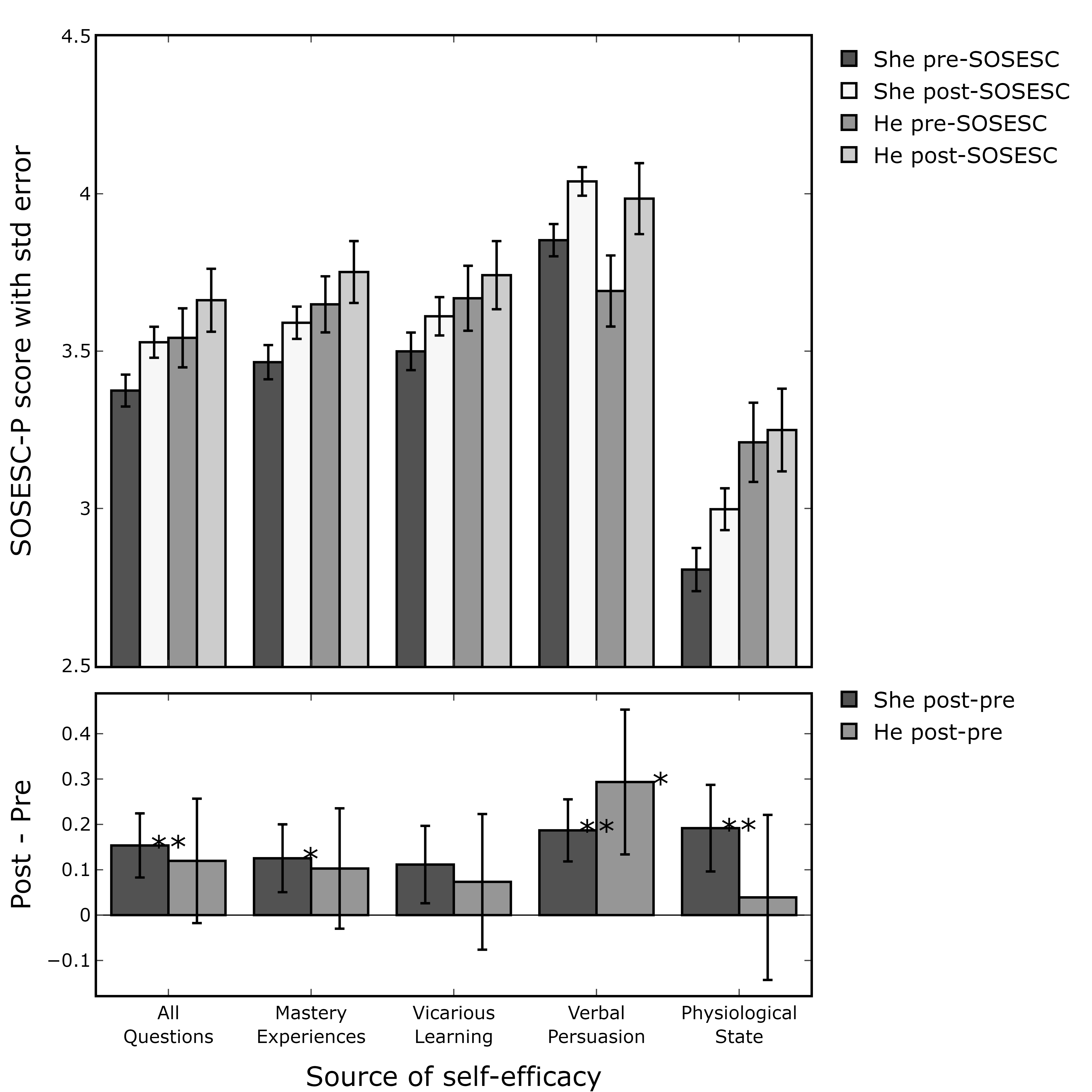}
	\caption{\label{fig:SE-source-results} Student self-efficacy as measured by the SOSESC-P survey disaggregated by pronoun identification from each of the sources identified by Bandura: mastery experiences, vicarious learning, verbal persuasion, and physiological state. The bottom portion of the plot shows the shift in each dimension. A ``*'' on the bottom plot indicates the effect is significant at the $p < 0.05$ level, ``**'' indicates a $p < 0.005$ significance.}
\end{figure}

For a more nuanced understanding, the top portion of figure~\ref{fig:SE-source-results} shows the pre- and post- score for each source of self-efficacy defined by Bandura broken down by pronoun self-identification. Error bars represent the standard error on the mean. The bottom portion, meanwhile, shows the shift from the beginning to the end of the semester. Note, in order to show more detail, the vertical axis on the top portion showing the self-efficacy scores is zoomed in on the range of 2.5 - 4.5, the interval containing all our data. All $p$-values thresholds are determined by a paired $t$-test. The bottom portion shows that the significant shift for she-identifying individuals overall is a manifestation of a positive shift across three of the four sources of self-efficacy: mastery experiences ($p < 0.05$), verbal persuasion ($p < 0.005$), and physiological state ($p < 0.005$). The only source of self-efficacy which did not show a significant increase for she-identifying students was in vicarious learning: a belief in success arising from watching others, including the instructor, be successful.

\subsection{\label{sec:results-se-aspects} Self-Efficacy in different attributes of the course}

\begin{figure}[htb!]
	\includegraphics[width=\columnwidth]{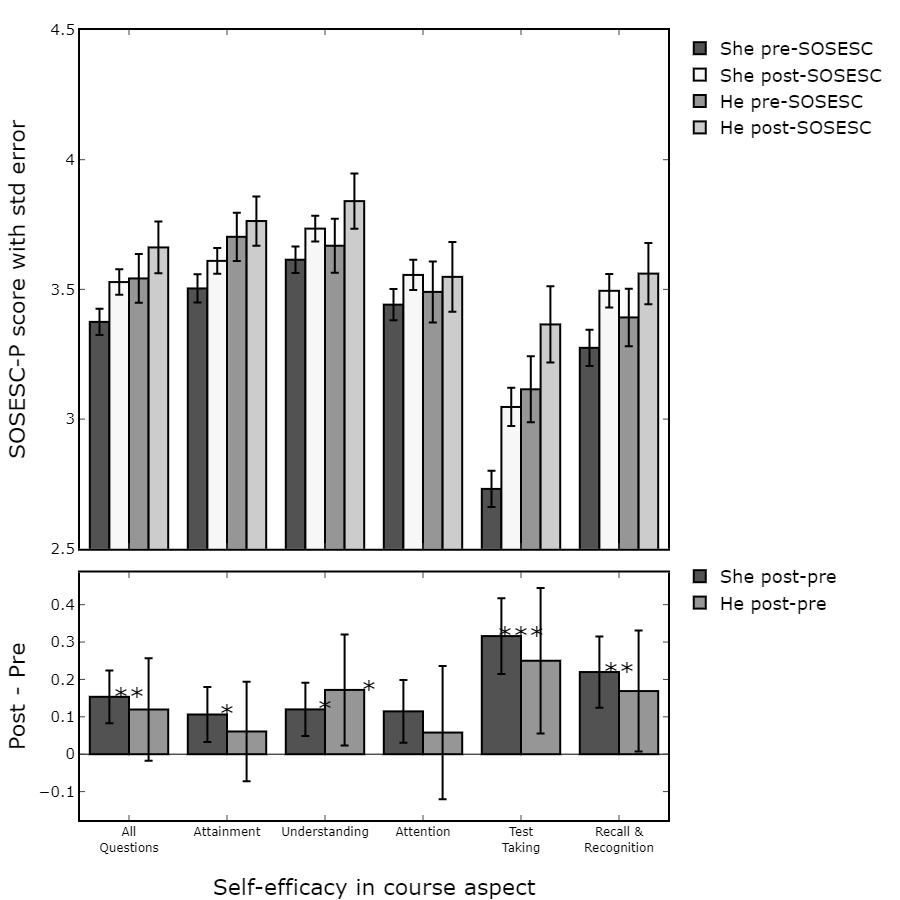}
	\caption{\label{fig:SE-attribute-results} Student self-efficacy as measured by the SOSESC-P survey disaggregated by pronoun identification ($N_{\mathrm{She}} = 91$, $N_{\mathrm{He}} = 37$) for the various attributes of the course: attainment, understanding, attention, test-taking, plus recall and recognition. The bottom portion of the plot shows the shift in each dimension. A ``*'' on the bottom plot indicates the effect is significant at the $p < 0.05$ level, ``**'' indicates a $p < 0.005$ significance, and ``***'' indicates $p< 0.0005$.}
\end{figure}

As described in section \ref{sec:SOSESC-P}, in addition to sources of self-efficacy, the SOSESC-P measures student self-efficacy from all sources for various attributes of a physics course including: attainment (getting good grades), understanding of content, ability to pay attention in class, test-taking, and recall and recognition. Our course’s SOSESC-P results for each of these attributes can be seen in Figure~\ref{fig:SE-attribute-results}. Again, the upper portion shows the pre and post scores disaggregated by pronoun, while the bottom portion shows the shift with $p$-value thresholds as determined by a paired $t$-test. As with the sources of self-efficacy, we see significant gains for she-identifying students in all attributes of the course except `attention.' A particularly large shift is visible in the area of `test taking,' with a $t = 4.392$, $p = 2 \times 10^{-5}$. For `he'-identifying individuals, limited statistics are again likely a factor. However, we still see a shift significant at the $p < 0.05$ level for this sub-population in the area of `understanding.'

\subsection{\label{sec:parity} Parity between she- and he-identifying individuals}

Due to the limited statistics for students who identify as `he', comparing the results of she- and he- identifying students is difficult. However, there are two dimensions where the differences were sufficient to be statistically significant: self-efficacy arising from physiological state and self-efficacy in the test-taking attribute of the course. These data are summarized in Table~\ref{tbl:parity}. The gap in the test taking attribute merits particular note. She-identifying individuals, on average, experienced their largest gains in this dimension. However, they only ``caught up'' to the starting value of he-identifying students who also experienced a gain in this area. This result is consistent with literature in mathematics education, where Arch suggests that she-identifying individuals tend to have more negative thoughts during exams and a lower value of their performance~\cite{Arch}. 

\begin{table}[tbh!]
\caption{\label{tbl:parity} A summary of two dimensions of self-efficacy where significant differences in end-of-semester SOSESC-P  results between she- and he-identified individuals can be observed. Errors are standard errors. }
\begin{tabular}{l l r r }
\hline \hline
                                                        &              & Physiological State & \hspace{2em} Test-Taking \\
\hline
\multirow{5}{*}{Pre \hspace{1.5em}}  & She         & $2.81 \pm 0.07$   & $2.73 \pm 0.07$ \\
                                                        & He           & $3.21 \pm 0.13$   & $3.11 \pm 0.13$ \\
                                                       & He - She   & $0.40  \pm 0.13$  & $ 0.38 \pm 0.13$ \\ 
                                                       & $t$           & $3.02$                & $2.82$              \\
                                                       & $p$          &  $0.003$              & $0.006$              \\
\hline 
\multirow{5}{*}{Post \hspace{1.5em}}  & She         & $3.00 \pm 0.06$   & $3.05 \pm 0.07$ \\
                                                       & He           & $3.25 \pm 0.13$   & $3.36 \pm 0.15$ \\
                                                       & He - She   & $0.25 \pm 0.12$  & $0.32 \pm 0.14$ \\ 
                                                       & $t$           & $1.88$                & $2.14$              \\
                                                       & $p$          &  $0.06$               & $0.03$              \\
\hline \hline
\end{tabular}
\end{table}

\section{\label{sec:discussion} Discussion}

To our knowledge, this is the first presentation of results which show a positive increase in self-efficacy for she-identifying people across multiple sources. Similarly, the she-identifying students’ positive shifts in self-efficacy across most measured attributes of the course seen in Figure~\ref{fig:SE-attribute-results} are promising, particularly those associated with `test taking.' We believe that specific features of the TBL pedagogy are important for these shifts.

The Sawtelle \textit{et al.} ~\cite{FIU-2010CP} and Dou \textit{et al.} ~\cite{FIU2016} results investigating self-efficacy in Modeling Instruction, are more typical for the literature. Sawtelle \textit{et al.}, using data from Fall 2008 – Fall 2009, saw significant negative shifts for standard lecture-based courses ($N = 175$). The only positive shift observed was for the reformed Modeling Instruction curriculum ($N = 70$), but only for self-efficacy arising from verbal persuasion and only for she-identifying individuals. When the sample size of students in Modeling Instruction was increased during the Falls of 2014 and 2015 to a total $N = 147$ in Dou et al, the result was a decrease in self-efficacy from all sources. Comparing these Modeling Instruction results with TBL directly is of limited value due to differences in class size and structure as well as student demographics. For example, Modeling Instruction is capped at 30 students, is calculus based, and has life-science and engineering majors together.  In comparison, TBL is 100 students, algebra-based, and overwhelmingly dominated with life-science students. However, the Modeling Instruction results are more typical of the results in the literature. 

\subsection{\label{sec:feedback} Immediate feedback as a source of improved self-efficacy}

\begin{figure}[tbh!]
	\includegraphics[width=\columnwidth]{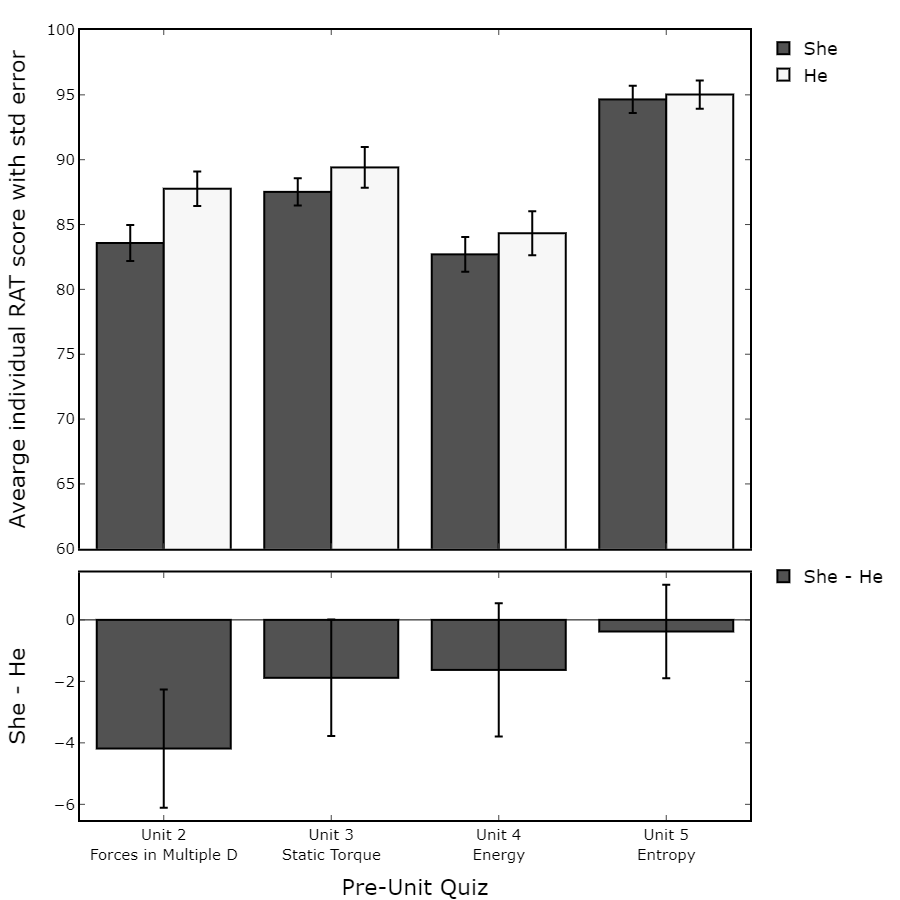}
	\caption{\label{fig:iQuizGrades} The average grade, with standard errors, on the individual portion of the 10-question readiness assessment test at the beginning of each unit for three sections of a first-semester IPLS TBL course $N = 206$. Note that the scores are rather high: a 78\% is the threshold for a B in this course.}
\end{figure}

\begin{figure}[tbh!]
	\includegraphics[width=\columnwidth]{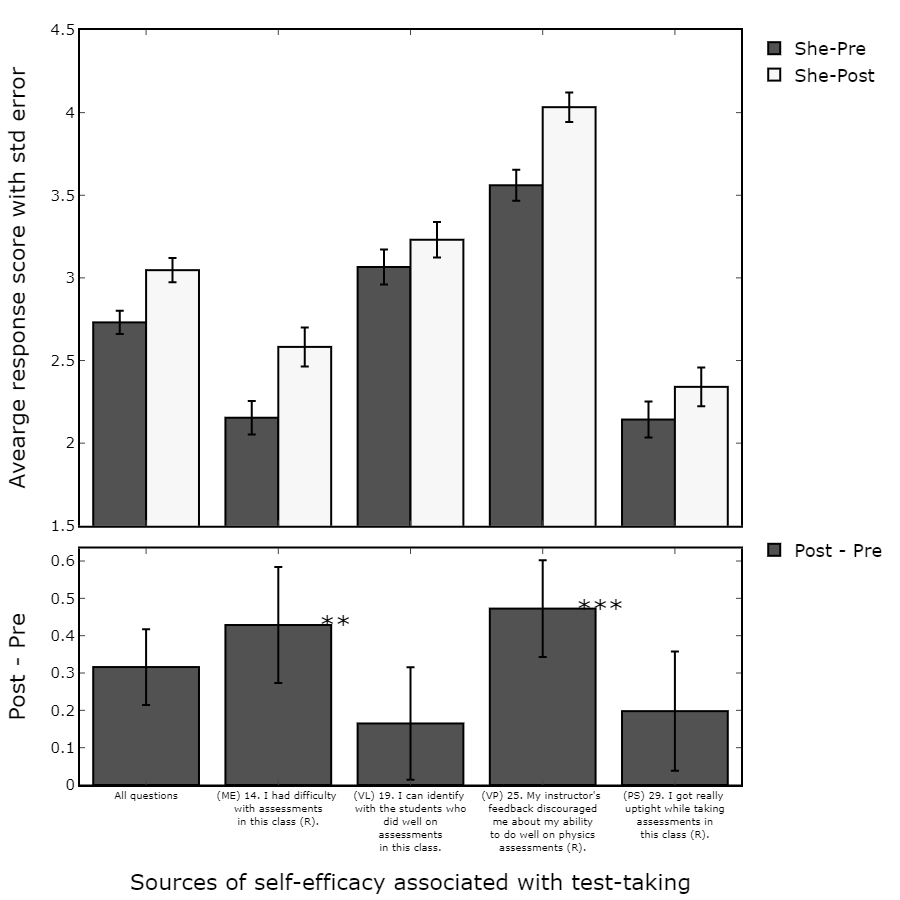}
	\caption{\label{fig:testTaking} The average responses to the four questions which specifically target self-efficacy connected with performance on exams and quizzes for she-identified students ($N = 91$). Each question is numbered and labeled by its associated source of self-efficacy: ME - mastery experience, VL - vicarious learning, VP - verbal persuasion, and PS - physiological state. The lower portion of the figure shows the pre-post shift with ``*'' indicating $p < 0.005$ and ``***'' indicating $p < 0.0005$.}
\end{figure}

The TBL pedagogy, like many active learning systems, has many opportunities for immediate feedback, which has been demonstrated to be important in student achievement~\cite{MakeItStick}. The flipped nature of the course and the subsequent Readiness Assessment Process described in section~\ref{sec:TBL}, is one facet specific to TBL which may help explain the self-efficacy improvements associated with mastery experience sources, particularly in the test taking attribute. The careful alignment between readings, preparatory homework, and the readiness assessment tests ensures that most students earn relatively high marks on the individual portions of the readiness assessment tests as shown in~\ref{fig:iQuizGrades}. Moreover, students know that they have been successful on the quiz immediately after they have completed it, as the team portion using the IF-AT cards provides immediate feedback~\cite{IFAT}. This immediate verification that they can learn physics, on their own, would seem to be a reasonably strong mastery experience source of self-efficacy. This belief in ability to execute the courses of action needed to do well on assessments may well then transfer to the exam context wherein students are required to solve problems different from what they have seen before. Looking at the specific breakdown of the $p < 0.0005$ shift in test taking shown in Figure~\ref{fig:testTaking} seems to support this supposition. The large shifts in self-efficacy associated with test taking for she-identifying students comes entirely from mastery experiences and from the verbal persuasion associated with the feedback. 

Immediate feedback is also an important feature of the primary in-class activity: working collaboratively at the whiteboards solving problems, constructing definitions, building concept maps, etc. with the support of the teaching team. As measured by observers using the SJSU RIOT~\cite{RIOT}, students at the beginning of a unit spend approximately one-quarter of class time at the whiteboards. In these first days, more scaffolding of problems and debriefing of solutions is required. The fraction of class time spent at the whiteboards then grows to one-half or higher by the end of the unit. This activity of working at the whiteboards provides another opportunity for immediate supportive feedback when students master a skill, providing sources of mastery experiences and verbal persuasion from the students' peers as well as the teaching team. As one student said in response to a question soliciting suggestions for improvements to the course on a reflection activity after the first exam, ``Being able to discuss as a group, critically problem solve, and then see what we did wrong all in the same day has been way more efficient in my learning because my original thought process is still there, and I can see immediately where I went wrong/right.''

\subsection{\label{sec:teams-SE} Team structure as a source of improved self-efficacy}

\begin{figure}[tbh!]
	\includegraphics[width=\columnwidth]{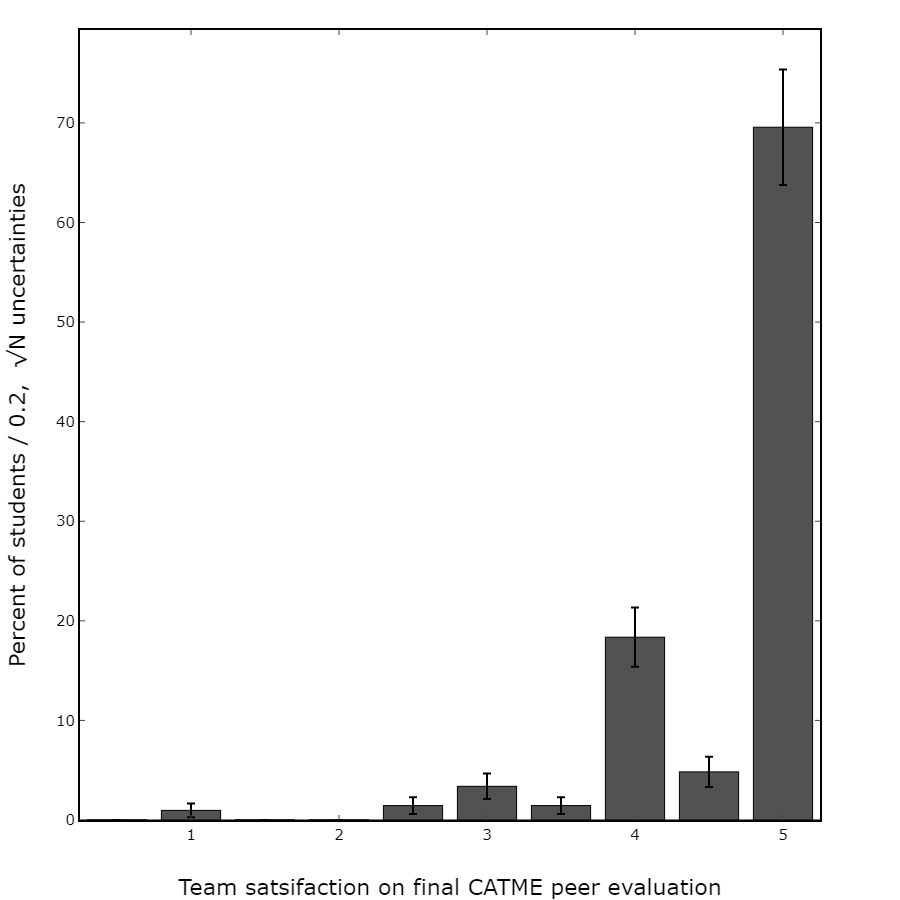}
	\caption{\label{fig:CATME-TS} Students' ($N = 206$) average response on a Likert scale to three questions specifically targeting team satisfaction given as part of the end-of-semester peer evaluation. A 5 represents ``Strongly agree.''}
\end{figure}

The comparatively large, intentionally diverse, and long-standing teams characteristic of TBL described in Section~\ref{sec:tbl-teams} may also be important for providing self-efficacy from verbal persuasion and physiological state, assuming student satisfaction with their teams. A portion of the end-of-semester peer evaluation process described in Section~\ref{sec:CATME}, asked students to respond on a Likert scale to three questions specifically targeting students' satisfaction with their teams: ``I am satisfied with my present teammates,'' ``I am pleased with the way my teammates and I work together,'' and ``I am very satisfied working in this team.'' The high average scores, with 5 representing ``Strongly Agree,'' shown in Figure~\ref{fig:CATME-TS} indicate that, in general, students were satisfied with their teams.

Student comments suggest a connection between the team structure and both verbal persuasion and physiological state. In the required end-of-semester peer evaluation, there is an opportunity for students to ``Please write your confidential comments to the instructor in the box below.'' One student commented, ``Physics is an intimidating class so, knowing that I was able to collaborate with peers helped with this predisposition.'' Another said, ``I think that everyone contributes different things to the team whether it is inside or outside of class, and whenever I am struggling to grasp a concept I can ask my team members who are more than willing to help me.'' Clearly these students are finding both physiological state and verbal persuasion sources of self-efficacy from their teams. Zeldin and Pajares suggest that people who identify as `she' will respond more significantly to verbal persuasion as a source of self-efficacy~\cite{Zeldin}. We therefore posit that these large, long-lasting teams may be contributors as to why we see significant shifts in self efficacy from verbal persuasion in she-identifying individuals in our study and suggest this as an area for future research.

\subsection{\label{sec:vl} Self-efficacy from vicarious learning}

\begin{figure}[tbh!]
	\includegraphics[width=\columnwidth]{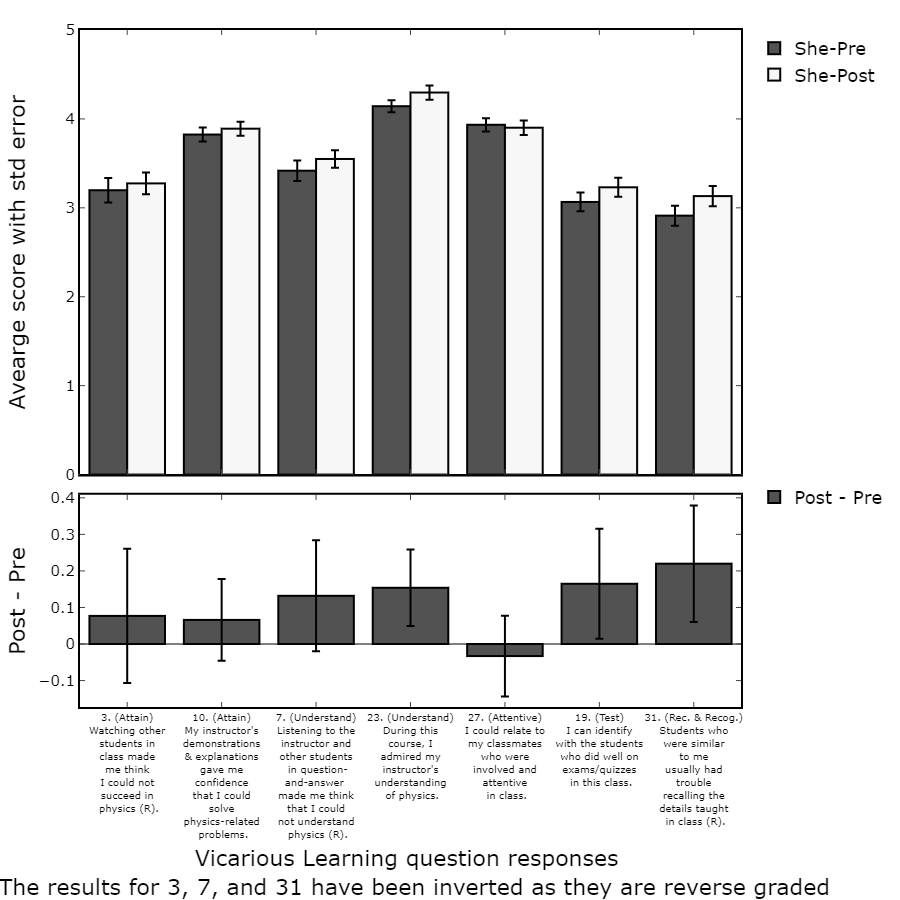}
	\caption{\label{fig:VL-questions} The pre- and post- results for each question relating to self-efficacy from vicarious learning sources for people who identify as `she' ($N = 91$) as well as the pre-/post- shift for each question. Questions 3, 7, and 31 are reverse graded. To simplify comparison, the results have been inverted: a response of `2' has been converted into a `4.'}
\end{figure}

The one source of self-efficacy in our study which does not exhibit a significant shift for people who identify as `she' is vicarious learning. The SOSESC-P questions examining this source of self-efficacy look at verbal persuasion from both the students' peers and from the instructor. Looking at the results for the individual questions in this category, Figure~\ref{fig:VL-questions} shows that none of the questions have a significant shift for `she'-identifying individuals at the $p < 0.05$ level using a paired $t$-test. This is particularly interesting in light of the work by Zeldin and Pajares which seems to suggest that vicarious learning may be an important source of self-efficacy for she-identifying students~ \cite{Zeldin}.

\section{\label{sec:results-log} Self-Efficacy as Predictor of Student Success} 

While we feel that improving student self-efficacy is an important course goal in-of-itself, we also feel that the relationship between self-efficacy and course-performance within the TBL environment merits investigation. After considering the particular features of the TBL pedagogy, we will define success in terms of individual exam scores and use a logarithmic regression to quantify the impact of self-efficacy over the other demographic factors explored in Section~\ref{sec:UMassIplsTbl}.

A logistic regression model follows the structure
$$ \ln \left[ \mathrm{odds} \left( \mathrm{score} > \mathrm{threshold} \right) \right] = b_0 + \sum_{k} b_k x_k$$ 
where $x_k$ are the various variables to be considered: pronoun, GPA, self-efficacy score, etc. Logistic regressions permit us to determine the odds that a student will pass some given threshold. For example, a logistic regression can predict the odds that a student will have an Individual Exam I score above 78\%. While a multiple linear regression would, in principle, permit us to predict a student's score as opposed to the odds of passing a threshold, we cannot strictly interpret the statistical significance of the results of a multiple linear regression due to the non-normalness of exam scores coupled with the ceiling effect which violate the strict assumptions of multiple linear regression. A logistical regression, on the other hand, permits statistical interpretation.

\subsection{\label{sec:success} Definitions of success and thresholds for model}

\begin{figure}[tbh!]
	\includegraphics[width=\columnwidth]{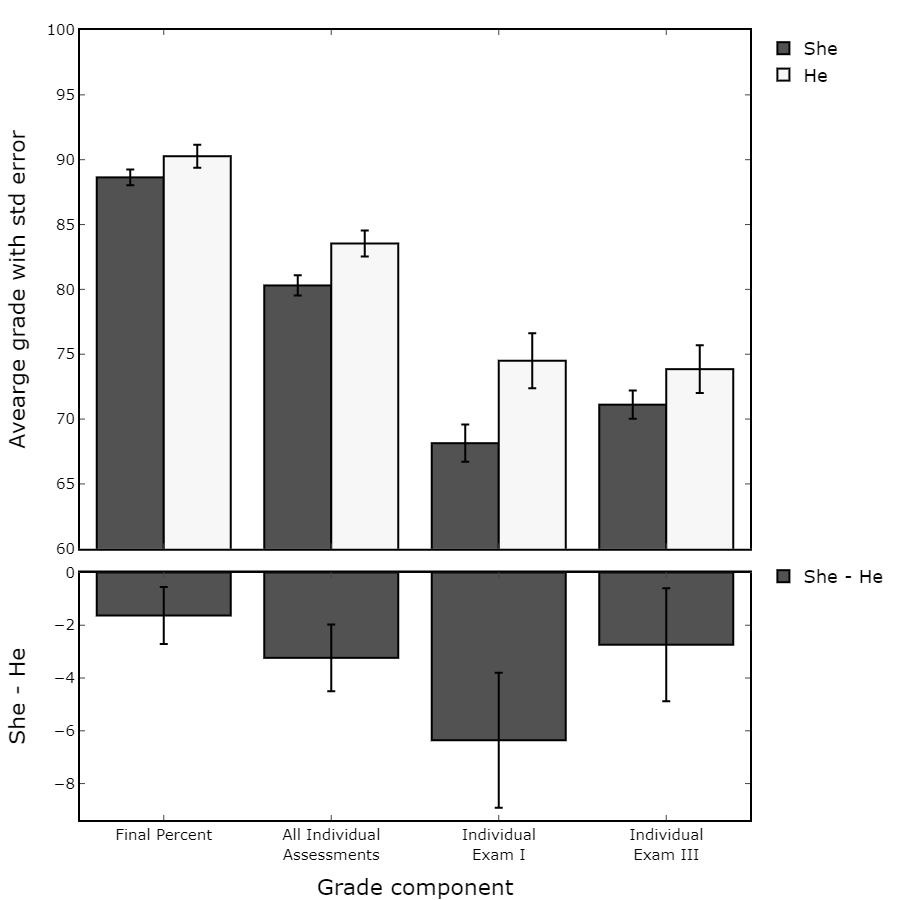}
	\caption{\label{fig:grades} The grades, dis-aggregated by pronoun identification ($N_{\mathrm{She}} = 91$, $N_{\mathrm{He}} = 37$), for the individual exams I and III as well as the average for all individual assessments, which includes beginning-of-unit quizzes. The total final grades are also shown for reference. The bottom portion of the plot shows the differential between she- and he- identifying students.} 
\end{figure}

In our analysis, logistic regressions will model three different metrics of success. The first metric is the final individual assessment grade combining all assessments students complete on their own: the individual portions of beginning-of-unit quizzes and exams. We will also look at the first and third individual exams (I and III) separately to look for changes between the beginning and end of the semester. Recall all exams in our course are cumulative and all are weighted equally. Exam I is after the units on kinematics/dynamics and comes at the 6-week mark. As discussed in section~\ref{sec:tbl-teams}, students are still in the process of forming teams at this point. Moreover, a significant amount of the material on individual Exam I would be covered in a typical high school physics class. Thus, we expect individual Exam I to be impacted by previous experience and less by the IPLS TBL pedagogy than later exams. Exam II adds static torque and conservation of energy across distance scales, while Exam III adds a statistical interpretation of entropy. We have chosen these criteria because we want to focus on the impact of self-efficacy on individual mastery of the material. We do not wish to reduce the importance of developing scientific collaborative skills in our course goals. However, individual exam performance is a very standard metric of interest to many parties throughout our institution. Figure~\ref{fig:grades} shows the grades for she- and he-identifying individuals for these three metrics as well as for the course as a whole. We see a significant gender gap in the average for individual Exam I which is reduced, but still present, by individual Exam III. As an additional note, the gap in individual total percent is lower than any exam, indicating that other, team-based, course components are compensating for these gaps, if incompletely.

For all measures of success, the minimum grade for a B on our grading scale, 78\%, will serve as the threshold for the logistic regression. Table~\ref{tbl:thresholds} shows the gap between she- and he-identifying individuals on the various metrics of success for the 78\% cutoff, as well as the two adjacent grade thresholds. Clearly, the three grade levels are statistically equivalent. From speaking to students, however, a grade of B seems to be a significant psychological step over a B-. In comparison, the distinction between B and B+ seems less important. Thus, we choose B as our threshold of success.

\begin{table}[tbh!]
\caption{\label{tbl:thresholds} The difference in the percent of she-identifying individuals versus he-identifying with a score less than the given threshold on each of the three individual measures of success.}
\begin{tabularx}{\columnwidth}{l r r r}
\hline \hline
 & \multicolumn{3}{l}{(\% She)  - (\% He) \textit{less} than threshold} \\
                 & \hspace{1em} Individual & \hspace{1em} Individual & \hspace{1em} All Individual \\
 Threshold &                            Exam I &                         Exam III &                       assessments \\
 \hline
 73\% (B-)  & $25.2 \pm 10.1$  & $15.2 \pm 10.0$  & $5.6 \pm 5.5$ \\
 78\% (B)   & $25.8 \pm 11.2$ 	& $10.4 \pm 11.7$ & $14.0 \pm 8.2$ \\
 81\% (B+) & $25.0 \pm 11.6$ 	& $13.9 \pm 12.4$ & $19.0 \pm 9.1$ \\
 \hline \hline
\end{tabularx}
\end{table}

\subsection{\label{sec:regression-results} Results of regression and discussion}

The results of our regression models are in Table~\ref{tbl:log-results} which compares the impact of a single self-efficacy measurement relative to demographic factors for pronoun preference and incoming GPA. Pronoun preference was included in the model by the binary \texttt{isFemale} variable which is 0 for he-identifying individuals and 1 otherwise. In other words, the three individuals in our study who identify with neither `she' nor `he' are aggregated with the 91 `she'-identifying students. We also investigated the impact of student major, but in no model did major add any predictive power.

\begin{table*}[tbh!]
\caption{\label{tbl:log-results} The results of our regressions predicting the odds of a student earning at least a B (78\%) on each of the three definitions of success: the first and third individual exams (individual exam I and individual exam III) as well as for all individual assessments which includes all exams and all beginning-of-unit quizzes. The models compare the impact of a single self-efficacy measurement from the SOSESC-P relative to pronoun preference and incoming GPA.}
\begin{tabular}{l r r r r r}
\hline \hline 
                &                                               & \hspace{0.5em} Demographics & \multicolumn{3}{c}{Impact of SOSESC-P Results} \\
                & coefficient $b_k$  & only                                     &\hspace{2em} pre-score & \hspace{2em} post-score & \hspace{1em} pre/post shift \\
\hline
\multirow{5}{*}{individual exam I}                      & Intercept				 & $-1.2 \pm 1.6$  & $-2.0 \pm 2.2 $             & $-6.0 \pm 2.2$  & $-0.7 \pm 1.6$ \\
                                                           & isFemale           & $-1.0 \pm 0.4$  & $-0.9 \pm 0.4$              & $-0.9 \pm 0.4$  & $-1.1 \pm 0.4$ \\
                                                           & GPA                 & $+0.4 \pm 0.4$  & $+0.4 \pm 0.4$            & $+0.2 \pm 0.1$ & $+0.2 \pm 0.4$ \\
                                                           & Self-efficacy		 & --                     & $+0.2 \pm 0.4$             & $+1.2 \pm 0.4$ & $+1.5 \pm 0.5$ \\
                                                           & LLR $p$-value   & $2.3\times10^{-2}$ & $5.2\times10^{-2}$ & $6.1\times10^{-4}$ & $4.7\times10^{-4}$ \\
\hline
\multirow{5}{*}{individual exam III}                     & Intercept			 & $-3.1 \pm 1.9$  & $-4.1 \pm 2.4$             & $-6.9 \pm 2.4$  & $-2.6 \pm 1.9$ \\
                                                             & isFemale         & $-0.7 \pm 0.4$  & $-0.6 \pm 0.4$             & $-0.5 \pm 0.4	$  & $-0.8 \pm 0.4$ \\
                                                             & GPA               & $+0.8 \pm 0.5$  & $+0.9 \pm 0.5$           & $+0.7 \pm 0.5$ & $+0.7 \pm 0.5$ \\
                                                             & Self-efficacy		 & --                     & $+0.3 \pm 0.4$            & $+1.1 \pm 0.4$ & $+1.2 \pm 0.5$ \\
                                                             & LLR $p$-value   & $3.2\times10^{-2}$ & $5.9\times10^{-2}$ & $1.2\times10^{-3}$ & $2.8\times10^{-3}$ \\
\hline
\multirow{4}{*}{Individual Assessments} & Intercept			 & $-4.4 \pm 2.1$  & $-4.4 \pm 2.6$             & $-9.3 \pm 2.8$  & $-4.0 \pm 2.1$ \\
                                                             & isFemale         & $-0.6 \pm 0.5$  & $-0.6 \pm 0.5$             & $-0.5 \pm 0.5$  & $-0.8 \pm 0.5$ \\
                                                             & GPA               & $+1.6 \pm 0.6$  & $+1.6 \pm 0.6$           & $+1.6 \pm 0.6$ & $+1.5 \pm 0.6$ \\
                                                             & Self-efficacy		 & --                     & $0.0 \pm 0.4$            & $+1.3 \pm 0.5$ & $+1.9 \pm 0.5$ \\
                                                             & LLR $p$-value & $6.5\times10^{-4}$ & $2.1\times10^{-3}$ & $1.8\times10^{-5}$ & $2.7\times10^{-6}$ \\
\hline \hline   
\end{tabular}
\end{table*}

For all three metrics of success, individual Exam I, individual Exam III, and total individual assessment average, we see that the incoming self-efficacy as measured by the SOSESC-P adds little predictive power relative to incoming demographics. However, the end-of-semester measurements and pre/post shifts are much stronger predictors. Even for the first assessment, individual Exam I, the incoming self-efficacy is not a strong predictor: the relevant coefficient $b_k = 0.2 \pm 0.4$ is consistent with zero. In contrast, self-efficacy scores which are dependent upon end-of-semester measurements, which take place seven weeks after individual Exam I, curiously seem to add better predictive power: $b_k = 1.2 \pm 0.4$. Perhaps, we are seeing reflected a difference between students with a growth versus fixed mindset as described in~\cite{Hochanadel}. Clearly, this is an interesting area for future research. 

By individual Exam III and the individual assessment average, self-efficacy is a more important predictor of success than students' pronoun identification. Consider the most predictive model based upon log-likelihood $p$-value which predicts individual assessment average using the shift in self-efficacy: $\mathrm{LLR} $p$-\mathrm{value} = 2.7\times10^{-6}$. In this model, the impact of students' pronoun preference has a coefficient of $-0.8 \pm 0.5$; the negative sign is indicative of the small pronoun gap still visible in Figure~\ref{fig:grades}. In comparison, the $b_k$ for the shift in self-efficacy is $+1.9 \pm 0.5$. Even though a firm cause-and-effect relationship cannot be definitively drawn from these results, clearly shifts in self-efficacy provide an important window into reducing performance gaps. While we consider improvements in self-efficacy to be important course goals in-of-themselves, further research should explore this connection.

\section{\label{sec:conclusions} Conclusions and impacts for instruction} 

We investigated the impact of the combination of a team-based pedagogy and an IPLS curriculum on students' self-efficacy. Improvement in self-efficacy is observed from all sources and across all measured attributes of a physics course for she-identifying students. The result was statistical parity between `she' and `he' identifying students by the end of the course on all dimensions except for those associated with test taking. In this dimension, we saw those who identify as `she' experience some of their largest gains, but were unable to ``catch up'' to `he'-identifying students. We postulate that these improvements are potentially driven by the large, diverse, long-standing teams distinctive of the team-based-learning pedagogy, along with the multiple opportunities for immediate feedback that this teaching system provides. 

We feel that improvements in self-efficacy are intrinsically important educational goals. However, we also find that shifts in self-efficacy are important predictors of student success on individual exams. In fact, the shift in self-efficacy is more important than a student's pronouns in predicting attaining at least a B on individual assignments. In this context, the asymmetric gains in self-efficacy between `he' and `she' identifying individuals contribute to the reduction in gender gap as the semester progresses. Future work could follow~\cite{Identifying-SE-MI} and focus on effective ways of further improving self-efficacy in the TBL environment.

\begin{acknowledgments}

The authors wish to acknowledge the support of the University of Massachusetts Amherst for their financial support of this work through various internal grants with particular thanks to the Center for Teaching and Learning. We particularly thank Brian Baldi for administering our surveys. We also acknowledge the useful conversations with Vashti Sawtelle regarding the SOSESC-P survey. 

\end{acknowledgments}

\appendix
\section{\label{app:problems} Sample homework problems}
As described in section~\ref{sec:TBL}, the preparatory homework problems focus on fundamental understanding. Below are two problems to illustrate.

\subsection{Problem from Unit I}
As described in Table~\ref{tbl:units}, the homework for the first unit introduces the concepts needed for kinematics and dynamics in one-dimension. This particular problem, which would be classified under the \textit{Remember} cognitive process dimension under the revised Bloom's Taxonomy of Krathwol~\cite{RevisedBlooms}, helps ensure students are clear on the subtleties of physics terminology:
\begin{quote}
\textit{Velocity} differs from \textit{speed} in that \textit{velocity} indicates a particle's \rule{1cm}{0.15mm} of motion. 
\textbf{Select the letter from the list that best complete the sentence}
\begin{enumerate}[label=\alph*)]
\item{position}
\item{direction}
\item{displacement}
\item{coordinates}
\item{acceleration}
\item{distance}
\end{enumerate}
\end{quote}

\subsection{Problem from Unit IV}
This unit focuses on energy. In this problem, which would be classified as \textit{Understand} under the revised Bloom's Taxonomy of Krathwol~\cite{RevisedBlooms}, students must classify forces as to whether or not they have an associated potential energy (are conservative): 
\begin{quote}
Check all of the forces below which do NOT have an associated potential energy.
\begin{itemize}
 \renewcommand{\labelitemi}{$\square$} 
 \item{Tension}
\item{Spring forces}
\item{Gravity}
\item{Friction}
\end{itemize}
\end{quote}

\bibliography{P131-2018-SE}

\end{document}